\documentclass[epj]{svjour}
\usepackage[utf8]{inputenc}
\usepackage{amsmath}
\usepackage{amssymb}
\usepackage{MnSymbol}
\usepackage{bbold}
\usepackage{graphics}
\usepackage[dvipsnames,usenames]{xcolor}
\usepackage[qm,braket]{qcircuit}

\begin{document}
\title{Faster spectral density calculation using energy moments}
\author{Jeremy Hartse\inst{1} \and Alessandro Roggero\inst{1,2,3}
}         

%
%
\institute{InQubator for Quantum Simulation (IQuS), Department of Physics, University of Washington, Seattle, WA 98195, USA \and Physics Department, University of Trento, Via Sommarive 14, I-38123 Trento, Italy \and INFN-TIFPA Trento Institute of Fundamental Physics and Applications,  Trento, Italy}

%
%
\abstract{
Accurate predictions of inclusive scattering cross sections in the linear response regime require efficient and controllable methods to calculate the spectral density in a strongly-correlated many-body system. In this work we reformulate the recently proposed Gaussian Integral Transform technique in terms of Fourier moments of the system Hamiltonian which can be computed efficiently on a quantum computer. One of the main advantages of this framework is that it allows for an important reduction of the computational cost by exploiting previous knowledge about the energy moments of the spectral density. For a simple model of medium mass nucleus like $^{40}$Ca and target energy resolution of $1$ MeV we find an expected speed-up of $\approx125$ times for the calculation of the giant dipole response and of $\approx50$ times for the simulation of quasi-elastic electron scattering at typical momentum transfers.}

%
\maketitle
\section{Introduction}

Response functions describe the linear response of a many-body system after an excitation and contain the same information as an inclusive reaction cross section. They can be expressed in terms of the spectral density operator $\delta(\hat{H}-\omega)$ of the Hamiltonian $\hat{H}$ describing the many-body system. First principle calculations of response function is in general extremely challenging for strongly correlated systems. A very powerful approach to study dynamical properties of many-body systems is to employ integral transform techniques which map the local spectral density into more manageable ground state expectation value which can be used to infer properties of the response function. This is the approach used in the Lorentz Integral Transform (LIT) method~\cite{Efros_1994,Efros_2007} and the more recent Gaussian Integral Transform (GIT) method~\cite{git2020,git2022}. Thanks to the ability to efficiently simulate the real time dynamics of many-body systems, simulations employing quantum computers offer the possibility to tackle the calculation of scattering cross sections from first principles (see e.g.~\cite{Klco_2022} for a recent review). Interestingly, in order to describe inclusive scattering in the linear response regime using the response function, integral transform techniques are also useful in the design of efficient quantum algorithms~\cite{Roggero2019,git2020}. It is therefore important and timely to extend the available techniques in order to reduce as much as possible the computational resources required to calculate the nuclear response function on a quantum device.
 
Given an Hamiltonian $\hat{H}$, an initial state $\ket{\Psi_0}$ and a Hermitian excitation operator $\hat{O}$, our goal is to evaluate the frequency dependent response function defined as
\begin{equation}
\begin{split}
S(\omega) &= \langle\Psi_0\lvert\hat{O}^\dagger\delta\left(\omega-\hat{H}\right)\hat{O}\rvert\Psi_0\rangle\\
&= \sum_m \left|\langle\Psi_0\lvert\hat{O}\rvert 
 m\rangle\right|^2 \delta\left(\omega-E_n\right)\\
\end{split}
\end{equation}
where in the last line we used the expansion on the eigenbasis $\{\ket{m}\}$ of $\hat{H}$. In general, it is difficult to directly use this definition as it requires a complete knowledge of the full energy eigenspectrum. 
The main idea is, similarly to the LIT and the GIT methods, to consider instead an integral transform with kernel $K$ given by
\begin{equation}
\begin{split}
\label{eq:phi_nu_def}
\Phi(\nu) &= \int d\omega K(\nu,\omega) S(\omega)\\
&= \langle\Psi_0\lvert\hat{O}^\dagger K\left(\nu,\hat{H}\right)\hat{O}\rvert\Psi_0\rangle\\
&=\sum_n \left|\langle\Psi_0\lvert\hat{O}\rvert n\rangle\right|^2 K\left(\nu,E_n\right)\;.
\end{split}
\end{equation}
Note that both the original transform and it's integral transform have units on inverse energy. Furthermore, in this work we will focus on translationally invariant kernel functions for which $K(\nu,\omega)=K(\nu-\omega)$, but extensions to the more general case are straightforward.

Once $\Phi(\nu)$ has been obtained, one usually attempts an inversion of the integral transform in order to obtain $S(\omega)$ back, however this procedure can introduce uncontrollable errors whenever the kernel function has compact support~\cite{Gl_ckle_2009,Barnea_2010}. A variety of approximate inversion techniques that introduce, more or less explicitly, additional smoothing to reduce these artifacts have been proposed in the past~\cite{Silver1990,Vitali2010,Burnier2013,Kades2020,Raghavan2021}. However, if the kernel function is chosen appropriately this last step might not be necessary~\cite{git2020,git2022}. In particular, it is convenient to consider kernels such that
\begin{equation}
\label{eq:sigma_approx}
\int_{-\Delta}^{\Delta} d\nu K(\nu)\geq1-\Sigma\;,
\end{equation}
A kernel satisfying Eq.~\eqref{eq:sigma_approx} is called $\Sigma${\it -approximate} with {\it resolution} $\Delta$~\cite{git2020}. The reason for this definition is that, in the commonly encountered situation where one is interested in observables of the form
\begin{equation}
Q(S,f) = \int d\omega S(\omega)f(\omega)\;,
\end{equation}
for some bounded function $f$, one can use directly the integral transform $\Phi$ to approximate $Q(S,f)$ with $Q(\Phi,f)$ with controllable error~\cite{git2020}. More intuitively, we can think of $\Phi$ as a finite width approximation of the original response function with an energy resolution given by $\Delta$ and additional tails controlled by the parameter $\Sigma$. The original response is then recovered in the limit $\Delta,\Sigma\to0$.

This approach was recently used in Ref.~\cite{git2022} to provide a controllable approximation of the spectral density using histograms derived from the integral response (see also~\cite{Sobczyk2022} for a recent application for computing spectral functions in light nuclei using the Coupled Cluster method). 

A direct procedure to construct integral kernels satisfying Eq.~\eqref{eq:sigma_approx} is to use a polynomial expansion such as
\begin{equation}
\label{eq:kernel_poly}
K(\nu-\omega) = \sum_n^\infty c_n(\nu) \phi_n(\omega)\;,
\end{equation}
with $\{\phi_n\}$ an orthonormal basis of polynomials. This leads directly to an alternative expression for the integral transform in this basis
\begin{equation}
\label{eq:phi_expansion}
\Phi(\nu) = \sum_n^\infty c_n(\nu)\langle\Psi_0\lvert\hat{O}^\dagger \phi_n\left(\hat{H}\right)\hat{O}\rvert\Psi_0\rangle = \sum_n^\infty c_n(\nu)m_n\;,
\end{equation}
where $m_n$ denotes the frequency moments s of the response function over these polynomials
\begin{equation}
\label{eq:moments}
m_n = \int d\omega \phi_n(\omega) S(\omega)\;.
\end{equation}
If the series in Eq.~\eqref{eq:phi_expansion} converges rapidly, one can then estimate the integral transform $\Phi(\nu)$ with a small error by keeping only the first $N$ terms in the expansion. Since the response function, and thus its integral transform, have units of inverse energy we will quantify the error in its approximation as $\varepsilon/\Omega$ with $\varepsilon>0$ and $\Omega$ a suitable energy constant. As shown in~\cite{git2020}, we can approximate a response function with a $\Sigma$-accurate kernel with resolution $\Delta$ using a basis of Chebyshev polynomials and a number of terms given by 
\begin{equation}
N\approx\widetilde{\mathcal{O}}\left(\frac{\|\hat{H}\|}{\Delta}\sqrt{\log\left(\frac{1}{\Sigma}\right)\log\left(\frac{\Omega}{\Delta\varepsilon}\right)}\right)\;,
\end{equation}
with $\|\hat{H}\|$ as the spectral norm of the Hamiltonian. The notation used here neglects subleading logarithmic factors, in particular we define $\widetilde{\mathcal{O}}(f(x))=\mathcal{O}(f(x)\log(f(x)))$ in line with previously reported estimates~\cite{git2020}.

This results applies to any possible response function and is helpful whenever the Chebyshev moments of the Hamiltonian can be computed efficiently. This can be done in practice with quantum computers by means of a suitable quantum walk~\cite{Childs2017,Low2017,Low2019,gilyen2019quantum,Subramanian2019} or in approximate way with classical methods, such as the Coupled Cluster approach~\cite{git2022,Sobczyk2022}. In many applications the spectral function has a number of distinctive features like being dominated by low energy contributions or displaying a distinct peak structure like e.g. in the quasi-elastic regime. Many of these features are captured by energy moments of the form~\cite{Orlandini_1991}
\begin{equation}
\label{eq:ene_mom}
\mu_n = \int d\omega \omega^n S(\omega)\;.
\end{equation}
For instance the qualitative shape of a quasi-elastic peak can be characterized with a good accuracy from the first few moments alone~\cite{ROSENFELDER1980188}. The advantage of using moments $\mu_n$ to characterize the spectral density is that in many situations they can be calculated explicitly using ground-state many-body methods (e.g. with Monte Carlo~\cite{RevModPhys.70.743,PhysRevLett.68.3682,RevModPhys.87.1067}). 

In this work we employ a Fourier basis for the polynomial expansion of the integral transform Eq.~\eqref{eq:phi_expansion} in order to incorporate this information into the calculation of the full spectral density and reduce the overall computational cost of the simulation. Thanks to their ability to efficiently simulate the real-time evolution of many-body systems, quantum computers are expected to be able to give us access to expectation values of the form
\begin{equation}
g_\psi(t) = \langle \psi\lvert e^{it\hat{H}}\rvert\psi\rangle\;,
\end{equation}
for states $\psi$ which can be easily prepared. The function $g_{\psi}(t)$ contains important information about the many body system, and several authors have proposed techniques to use this to get access to a variety of observables like excitation energies, the local spectral density and even thermodynamic expectation values~\cite{OBrien_2019,Somma_2019,Lu2021,Guzman2021,Cortes2022}.

Under general consideration we can expect two main time scales to play a role: first, in order to obtain a frequency resolution of order $\Delta$, the maximum time required will need to scale as $T=\mathcal{O}(1/\Delta)$; second, since the full energy spectrum is contained in a frequency interval of size $\|\hat{H}\|$, a time step $\delta t=\mathcal{O}(1/\|\hat{H}\|)$ will guarantee a perfect reconstruction without aliasing thanks to the Shannon-Nyquist theorem. The combination of these two time scales predicts a scaling of the number of time-steps required as $T/\delta t=\mathcal{O}(\|\hat{H}\|/\Delta)$. 
However, as shown recently in Ref.~\cite{Lu2021}, in situations where the spectral function has an energy variance $\sigma^2\ll\|\hat{H}\|^2$, a good Fourier approximation can be obtained with a scaling $T/\delta t=\mathcal{O}(\sigma/\Delta)$ instead. The goal of this work is to formalize this intuition and provide rigorous error bounds allowing use of prior information about the energy moments of the spectral function in order to reduce the number of expectation values required.

In Section~\ref{sec:f_reconstruction} we present the general framework employing Fourier moments to approximate the response function and in specialize the treatment to the Gaussian Integral Transform in Section~\ref{ssec:git_fourier}. Section~\ref{sec:examples} details two example response functions that are reconstructed with this method and the improved efficiency afforded by employing energy moments. In Section \ref{sec:concl} we conclude by discussing applications of this method to typical scattering properties in nuclear physics, and the estimated improvement in number of Fourier moments required.

\section{Fourier based reconstruction}
\label{sec:f_reconstruction}
In order to use directly a discrete Fourier transform to perform the expansion of the integral kernel in Eq.~\eqref{eq:kernel_poly} it's convenient to define a periodic extension of the kernel function as follows~\cite{Somma_2019}
\begin{equation}
\label{eq:period_kernel}
K^\chi(\nu,\omega) = \sum_{n=-\infty}^\infty K\left(\nu,\omega+n\chi\|\hat{H}\|\right)\;,
\end{equation}
where the period $P=\chi\|\hat{H}\|$ is expressed here as a function of the spectral norm. This convention allows us to truncate the sum at the first term if we take $\chi\gg1$: since the spectrum is bounded only the term with $n=0$ has a finite contribution. As $\chi$ is reduced more terms will start to contribute to the sum and the extended integral kernel $K^\chi$ will start to deviate from the original. The main advantage of this construction is that we can express directly the periodically-extended kernel as a Fourier series
\begin{equation}
\label{eq:kernel_chi_exp}
K^\chi(\nu,\omega) = \frac{1}{\chi\|\hat{H}\|} \sum_{n=-\infty}^\infty k^\chi_n\left(\nu\right) \exp\left(-i\frac{2\pi n}{\chi\|\hat{H}\|}\omega\right)\;,
\end{equation}
where the Fourier transformed coefficients are given by
\begin{equation}
k^\chi_n\left(\nu\right) = \int_{-\infty}^\infty d\omega  \exp\left(i\frac{2\pi n}{\chi\|\hat{H}\|}\omega\right) K(\nu,\omega)\;.
\end{equation}
The main advantage of working with $K^\chi$ and this Fourier series representation is that we can now express a general integral transform of the response function as a linear combination of expectation values of the time-evolution operator. From the definition in Eq.~\eqref{eq:phi_nu_def} we can in fact express the new integral transform as follows
\begin{equation}
\begin{split}
\Phi^\chi(\nu) &= \langle\Psi_0\lvert\hat{O}^\dagger K^\chi(\nu,\hat{H})\hat{O}\rvert\Psi_0\rangle\\
&=\frac{1}{\chi\|\hat{H}\|} \sum_{n=-\infty}^\infty k^\chi_n(\nu) \langle \Psi_0\lvert \hat{O}^\dagger e^{-in\delta t\hat{H}}\hat{O}\rvert\Psi_0\rangle
\end{split}
\end{equation}
with finite time-steps of size $\delta t=2\pi/(\chi\|\hat{H}\|)$. Provided we choose a smooth integral kernel in the $\omega$ variable, the series expansion converges quickly and we can therefore obtain an accurate approximation of the integral transform by taking a truncation to some finite order
\begin{equation}
\label{eq:phi_finite_N}
\Phi^\chi_N(\nu) = \frac{1}{\chi\|\hat{H}\|} \sum_{n=-N}^N k^\chi_n(\nu) \langle \Psi_0\lvert \hat{O}^\dagger e^{-in\delta t\hat{H}}\hat{O}\rvert\Psi_0\rangle\;.
\end{equation}
For example, if we consider integral kernels in $C^\infty$, the convergence will be in general super-polynomial in $N$ and,  importantly, rigorous bounds on the truncation error can be found with relatively straightforward calculations. It is important to note at this point that the use of a smoothing kernel with a finite energy resolution $\Delta$ is critical to ensure that the expansion is well-behaved: indeed a direct approximation of the response function $S(\omega)$ using a finite Fourier sum would have encountered difficulties due to the discreteness of the energy spectrum for a finite system, which creates sharp peaks. The difference between the present approach and related ones (from Refs.~\cite{git2020,git2022}) with more heuristic smoothing procedures like the Maximum Entropy Method~\cite{Silver1990,Burnier2013} or alternative polynomial expansion methods like the Kernel Polynomial Method~\cite{Wei_e_2006} is that the energy window over which the smoothing is applied is fully under control and all errors can be accounted for. The price to pay to be able to use higher energy resolution is a corresponding increase in the number of terms in the expansion Eq.~\eqref{eq:phi_finite_N} which corresponds to a comparable increase in the maximum time $T=N\delta t$ over which one needs to be able to simulate the many-body dynamics. We will discuss this in more detail for a specific choice of integral kernel in the next section. 

\subsection{Gaussian Integral Transform}
\label{ssec:git_fourier}

As shown already in Ref.~\cite{git2020}, the Gaussian Integral Transform (GIT) is particularly useful for the purpose of obtaining a $\Sigma$-accurate integral transform with resolution $\Delta$ with a fast converging polynomial expansion like Eq.~\eqref{eq:kernel_poly}. This result is rather intuitive since a Gaussian envelope is a perfect trade-off between good resolution in frequency and small widths in the time domain. The original construction from Ref.~\cite{git2020} used a Chebyshev expansion for the polynomial basis, however in this work we instead explore an expansion of the Gaussian kernel into Fourier modes.

The first step is to determine an appropriate value for the width $\Lambda$ of the Gaussian kernel in order to satisfy the condition in Eq.~\eqref{eq:sigma_approx}. A direct calculation gives
\begin{equation}
\frac{1}{\sqrt{2\pi}\Lambda}\int_{-\Delta}^{\Delta}d\nu\exp\left(-\frac{\nu^2}{2\Lambda^2}\right)=\text{erf}\left(\frac{\Delta}{\sqrt{2}\Lambda}\right)\geq1-\Sigma\;.
\end{equation}
Using the upper-bound on the complementary error function $\text{erfc}(x)=1-\text{erf(x)}$ by a Gaussian
\begin{equation}
\label{eq:erfc_bnd}
\text{erfc}(x) \leq \exp(-x^2)\;,
\end{equation}
we can find the following sufficient condition~\cite{git2020}
\begin{equation}
\label{eq:for_lambda}
\Sigma\geq \exp\left(-\frac{\Delta^2}{2\Lambda^2}\right)\quad\Rightarrow\quad\Lambda \leq \frac{\Delta}{\sqrt{2\log(1/\Sigma)}}\;.
\end{equation}

Following the notation from Eq.~\eqref{eq:period_kernel} and Eq.~\eqref{eq:kernel_chi_exp}, we can then express the periodically extended Gaussian kernel as 
\begin{equation}
G^\chi(\nu,\omega) = \frac{1}{\chi\|\hat{H}\|} \sum_{n=-\infty}^\infty g^\chi_n(\nu) \exp\left(-i\delta t n\omega\right)\;
\end{equation}
with the corresponding Fourier coefficients
\begin{equation}
\label{eq:fcoeff}
g^\chi_n(\nu) = \exp\left(-i\delta t n\nu\right)\exp\left(-\frac{\delta t^2\Lambda^2}{2} n^2\right)\;.
\end{equation}
In the expressions above we used the definition $\delta t=2\pi/(\chi\|\hat{H}\|)$ for the time-step. We can now write the approximate integral transform obtained by truncating the series as
\begin{equation}
\Phi^\chi_N(\nu) = \frac{1}{\chi\|\hat{H}\|} \sum_{n=-N}^N g^\chi_n(\nu) \langle \Psi_0\lvert \hat{O}^\dagger e^{-in\delta t\hat{H}}\hat{O}\rvert\Psi_0\rangle\;.
\end{equation}

The ideal integral transform $\Phi(\nu)$ would be obtained in principle by choosing $\Lambda$ in order to satisfy Eq.~\eqref{eq:for_lambda}, taking $\chi=1$ and letting $N\to\infty$. The error in the approximation $\Phi^\chi_N$ is caused only by taking a finite number of terms $N$ and (possibly) reducing the value of $\chi$ which parametrizes the frequency interval used in the periodic extension. For a fixed value of the frequency $\nu$ we can then bound the error with
\begin{equation}
\lvert\Phi^\chi_N(\nu)-\Phi(\nu)\rvert \leq \lvert\Phi^\chi(\nu)-\Phi(\nu)\rvert+\lvert\Phi^\chi_N(\nu)-\Phi^\chi(\nu)\rvert\;,
\end{equation}
where we simply used the triangle inequality. The advantage is that the two error contributions on the right hand side can be bounded individually in a simpler way. We will denote the first error on the right hand side as $\epsilon_P$ and the second as $\epsilon_N$. In addition to these two sources of error, we need also to account for the fact we will estimate the Fourier moment
\begin{equation}
m^\chi_n = \langle \Psi_0\lvert \hat{O}^\dagger e^{-in\delta t\hat{H}}\hat{O}\rvert\Psi_0\rangle
\end{equation}
by computing the expectation value with a finite statistical sample leading to an additional error 
\begin{equation}
\epsilon_S = \lvert\Phi^\chi_N(\nu)-\overline{\Phi}_N^\chi(\nu)\rvert\;,
\end{equation}
where we have denoted by $\overline{\Phi}_N^\chi(\nu)$ the finite sample estimate of $\Phi_N^\chi(\nu)$.
The total error will be given then by the sum of all contributions $\epsilon=\epsilon_P+\epsilon_N+\epsilon_S$.

Before presenting our results for the complexity of estimating an accurate approximation of the integral transform $\Phi(\nu)$ it is convenient to specify several of the conventions we use, note however that the results provided in App.~\ref{sec:error_bounds} are completely general and do not depend on these conventions. First of all, since the response function (and thus it's integral transform) have dimensions of inverse energy, we will consider a dimensionless error parameter $\varepsilon>0$ obtained using a suitable energy scale $\Omega$. That is, we want to find values $\chi$ and $N$ for which
\begin{equation}
\label{eq:error_metric}
\lvert\overline{\Phi}_N^\chi(\nu)-\Phi(\nu)\rvert \leq \frac{\varepsilon}{\Omega}\;.
\end{equation}
Second, we consider the situation where we are interested in approximating the response function on a finite energy window $\omega\in[\omega_{\min},\omega_{\max}]$. For many applications in nuclear physics is also customary to shift the Hamiltonian by the ground state energy so that all frequencies become positive. Since both of these considerations apply directly to the integral transform, we will consider the situation where $\hat{H}$ has been shifted so that the ground state is at $\omega_{\min}=-\|\hat{H}\|$ consider a range $[-\|\hat{H}\|,-\|\hat{H}\|+\delta\nu]$ for the energies we want the value of $\Phi(\nu)$. Finally, since we are interested in computing the Fourier moments $m^\chi_n$
on a quantum computer, we will consider the situation where the excitation operator has been appropriately rescaled so that the state $\hat{O}\rvert\Psi_0\rangle$ is normalized to one. This directly implies that the zeroth moment $\mu_0$ will also be equal to one. A rescaling of this form has been employed already in past works on quantum algorithms for the response function~\cite{Roggero2019,git2020,Roggero2020nu,git2022,Baroni2022} and is natural when using efficient methods for preparing the excited state (see~\cite{Roggero2020exc}).

The only error term that depends directly on the energy variance $\sigma^2=\mu_2-\mu_1^2$ is the first one while the other one will only depend parametrically on the chosen period $P=\chi\|\hat{H}\|$. We start with the truncation error which can be bounded easily using standard techniques resulting in the following requirement for the number of terms
\begin{equation}
\label{eq:n_bound_main}
N\geq \frac{\chi\|\hat{H}\|}{\sqrt{2\pi}\Lambda}\sqrt{\log\left(0.4\frac{\Omega}{\varepsilon_N\Lambda}\right)}\;.
\end{equation}
A full derivation of this result can be found in App.~\ref{sec:epsilon_N}. The dependence on the error is typical for Gaussian kernels (c.f.~\cite{git2020,git2022}). The statistical contribution of the error can be controlled, with confidence level $1-\delta$, by requiring in the worst case a number of experiments given by
\begin{equation}
\begin{split}
S&=N\frac{\Omega^2}{\varepsilon_S^2\Lambda^2}\log\left(\frac{2}{\delta}\right)\\
&=\frac{\chi\|\hat{H}\|\Omega^2}{\sqrt{2}\pi\varepsilon_S^2\Lambda^3}\sqrt{\log\left(0.4\frac{\Omega}{\varepsilon_N\Lambda}\right)}\log\left(\frac{2}{\delta}\right)\;,
\end{split}
\end{equation}
where in the second line we used the estimate from Eq.~\eqref{eq:n_bound_main}. We present a full derivation of this result in App.~\ref{sec:stat_err_app}. As we can see, the sample complexity of the scheme is directly proportional to the factor $\chi$ controlling the period.

The remaining error $\epsilon_P$, controlled by the choice of $\chi$, can be found under to separate situations: the general case where we do not use information from known energy moments $\mu_n$ and the typical case where we have information at least on the energy variance. As mentioned in the text following Eq.~\eqref{eq:period_kernel}, in the first situation we want to take $\chi\gg1$ in order to minimize the error. In particular, as shown explicitly in App.~\ref{ssec:general_chi}, we find that for $\epsilon_P\leq\varepsilon_P/\Omega$ the following choice would be sufficient
\begin{equation}
\label{eq:chi_gen_main}
\chi=2+\frac{\sqrt{2}\Lambda}{\|\hat{H}\|}\sqrt{\log\left(\frac{2\Omega}{\varepsilon_P\Lambda}\right)}\;.
\end{equation}
In the limit $\Lambda\to0$ we recover the Shannon-Nyquist theorem which gives $\chi=2$ for a perfect reconstruction. At this point we can look for the asymptotic scaling of both $N$ and $S$ while guaranteeing a total error $\varepsilon_P+\varepsilon_N+\varepsilon_S\leq\varepsilon$. If take $\varepsilon_P=\varepsilon_N=\varepsilon_S=\varepsilon/3$ we find immediately
\begin{equation}
\begin{split}
\label{eq:nbound_old}
N= &\widetilde{\mathcal{O}}\left(\frac{\|\hat{H}\|}{\Delta}\sqrt{\log\left(\frac{1}{\Sigma}\right)\log\left(\frac{\Omega}{\varepsilon\Delta}\right)}\right.\\
&\left.\quad\quad\quad\quad\quad\quad+\log\left(\frac{\Omega}{\varepsilon\Delta}\sqrt{\log\left(\frac{1}{\Sigma}\right)}\right)\right)\;,
\end{split}
\end{equation}
which, apart from a logarithmic correction, is the same result found for the Chebyshev version of the GIT protocol. In turn we find that the required number of samples will scale in general as
\begin{equation}
S=\widetilde{\mathcal{O}}\left(\frac{\|\hat{H}\|\Omega^2}{\varepsilon^2\Delta^3}\log^{3/2}\left(\frac{1}{\Sigma}\right)\log\left(\frac{1}{\delta}\right)\right)
\end{equation}
again similar to the Chebyshev version (c.f. App.~\ref{sec:stat_err_app} and the orginal work Ref.~\cite{git2020}). The main difference is that the Chebyshev implementation avoids the (necessary) approximation of the time-evolution operator. If simulations schemes with optimal asymptotic scaling with the error are used, like the one based on qubitization~\cite{Low2017,Low2019}, the Chebyshev GIT will have an advantage in terms of gate counts over its Fourier version described here.

We now turn to the main result of this work where instead we use information about first two energy moments to design a scheme with better scaling. The main result we use is the Chebyshev inequality which, in terms of response functions, states that (assuming $\mu_0=1$)
\begin{equation}
\begin{split}
\label{eq:Cheb_bound}
1-\int_{\mu_1-\Gamma}^{\mu_1+\Gamma}d\omega S(\omega)&=Prob\left[\left|\omega-\mu_1\right|\geq\Gamma\right]\leq\frac{\sigma^2}{\Gamma^2}\;,
\end{split}
\end{equation}
for any positive constant energy $\Gamma$. The idea is to use Eq.~\eqref{eq:Cheb_bound} to constrain the integrated strength of the spectrum away from the mean. The interested reader can find the full derivation in App.~\ref{sec:pwithmom}, the final result is
\begin{equation}
\chi=  \frac{2.7}{\varepsilon_P^{1/3}} \frac{\Omega^{1/3}\sigma^{2/3}}{\|\hat{H}\|}+\frac{\delta\nu}{\|\hat{H}\|}\;,
\end{equation}
which is valid for $\Lambda\leq2\sigma$ (see App.~\ref{sec:pwithmom} for an estimate valid also in a lower resolution regime). This immediately gives an estimate for the required number of moments
\begin{equation}
\begin{split}
\label{eq:nmom_var}
N=&\mathcal{O}\left(\frac{1}{\Delta}\left(\frac{\Omega^{1/3}\sigma^{2/3}}{\varepsilon^{1/3}}+\delta\nu\right)\sqrt{\log\left(\frac{1}{\Sigma}\right)\log\left(\frac{\Omega}{\varepsilon\Delta}\right)}\right)\\
\end{split}
\end{equation}
Several comments are in order at this point. First of all, the scaling of $N$ with the target error $\varepsilon$ is exponentially worse than the original result Eq.~\eqref{eq:nbound_old} which didn't use energy moments. For situations that require very small errors $\varepsilon\lesssim \Omega\sigma^2/\|\hat{H}\|^3$ then the general scheme with $\chi$ from Eq.~\eqref{eq:chi_gen_main} should be used instead. This drawback can be mitigated if more information is available. For instance if we know the value of the $n$ central moment $\widetilde{\mu_n}$ then we can bring the cost down to
\begin{equation}
\begin{split}
\label{eq:nmom_high}
N=&\mathcal{O}\left(\frac{\Omega^{1/(n+1)}\widetilde{\mu_n}^{1/(n+1)}}{\Delta\varepsilon^{1/(n+1)}}\sqrt{\log\left(\frac{1}{\Sigma}\right)\log\left(\frac{\Omega}{\varepsilon\Delta}\right)}\right.\\
&\left.\quad\quad\quad+\frac{\delta\nu}{\Delta}\sqrt{\log\left(\frac{1}{\Sigma}\right)\log\left(\frac{\Omega}{\varepsilon\Delta}\right)}\right)\;.
\end{split}
\end{equation}
See App.~\ref{sec:pwithmom} for additional details and a proof.

The second important comment we can make about the result Eq.~\eqref{eq:nmom_var}, which also applies to its generalization Eq.~\eqref{eq:nmom_high}, is that if we want to reconstruct the energy spectrum over a large frequency range scaling as $\mathcal{O}(\|\hat{H}\|)$ then the second contribution proportional to $\delta\nu$ will dominate and we recover again the general result obtained without moments. This somewhat counter-intuitive behavior can be understood by noticing that our error metric in Eq.~\eqref{eq:error_metric} measures the error in a pointwise fashion: a very narrow but tall peak in the energy spectrum will contribute to the error even if its spectral weight is small. For specific applications, like the calculation of energy histograms presented in Ref.~\cite{git2022}, better bounds could be obtained. We leave this development for future work.

Finally we note that all the results presented in this work depend on an arbitrary energy scale $\Omega$ which controls the energy dimension of the final error. An appropriate choice for this parameter depends on the specific physical application desired. For example, in applications where one is interested in obtaining histograms of the spectrum like in Ref.~\cite{git2022}, the choice $\Omega=\Delta$ seems appropriate. 

\section{Numerical examples}
\label{sec:examples}

In this section we present some numerical experiments that show in practice the savings afforded by the method present here. In order to show the general trend, we will employ two simple model response functions $S_A(\omega)$ and $S_B(\omega)$. In both cases we produce a peak at low frequencies according to a skewed Gaussian distribution
\begin{equation}
    S_{peak}(\omega) = \frac{1}{\beta\sqrt{2\pi}}\exp{\frac{-(\omega-\xi)^2}{2\beta^2}}\left(1+\text{erf}\left( \frac{\alpha(\omega - \xi)}{\sqrt{2}\beta}\right) \right)\;,
\end{equation}
where $ \xi$ is the location of the peak, $ \beta$ is the scale of the distribution and $ \alpha$ the skewedness. In order to explore the impact of a tail at high energies we also use 
\begin{equation}
S_{tail}(\omega) = \left\{\begin{matrix}
0&\text{for}\;\;\omega<\omega_{thr}\\
\lambda\rho\left(\left|\omega-\omega_{thr}\right|^\gamma+\rho\right)^{-1}&\text{for}\;\;\omega\geq\omega_{thr}    
\end{matrix}\right.
\end{equation}
Here $\lambda$ is an overall normalization, $\rho$ sets the scale and $\gamma$ the exponent of a power-law decaying tail.

\begin{figure}[t]
  \resizebox{0.49\textwidth}{!}{ \includegraphics{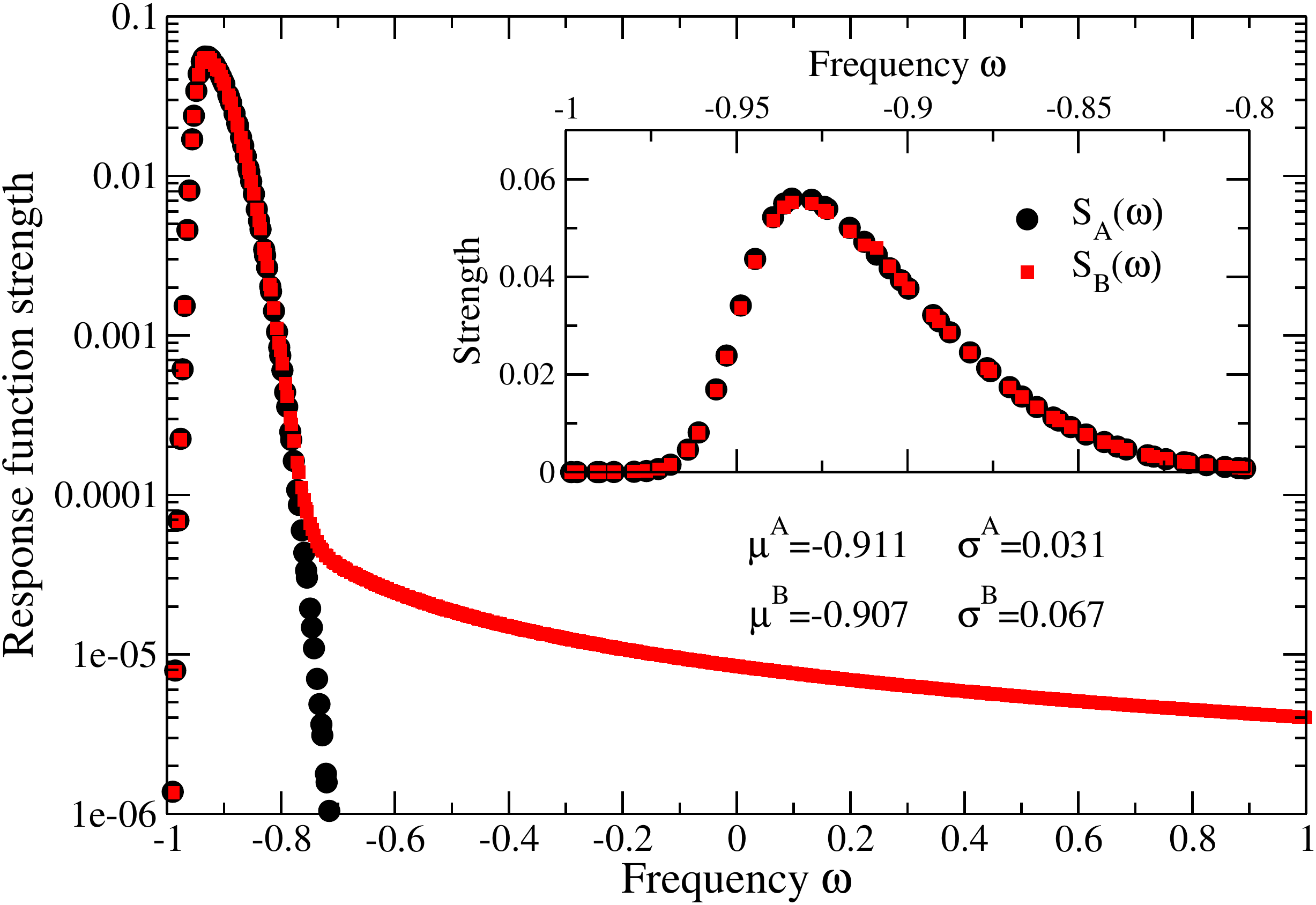} }
  \caption{The two model response functions shown as a function of frequency: black dots are for $S_A(\omega)$ while red squares for $S_B(\omega)$. The inset shows the same responses but on a narrower range and with linear scale. Also indicated are the mean $\mu$ and the square root of the variance $\sigma$ for both distributions. }
  \label{fig:models}
\end{figure}

In the numerical tests shown here we took $\|\hat{H}\|=1$ and $512$ eigenvalues $\{\omega_k\}$ distributed over the whole spectrum. The two model response functions we consider are
\begin{equation}
\begin{split}
S_A(\omega) &= \sum_{k} \frac{S_{peak}(\omega_k)}{\sum_k S_{peak}(\omega_k)}\delta(\omega-\omega_k)\\
S_B(\omega) &= \sum_{k} \frac{S_{peak}(\omega_k)+S_{tail}(\omega_k)}{\sum_k\left(S_{peak}(\omega_k)+S_{tail}(\omega_k)\right)}\delta(\omega-\omega_k)
\end{split}
\end{equation}
where the normalization in the denominators is chosen in order to guarantee that $\mu_0=1$ for both response functions. The parameters for the response functions are
\begin{equation}
\xi=\omega_{thr}=-0.95\;\;\alpha=5\;\;\beta=0.05\;\;\lambda=1\;\;\rho=0.002\;.
\end{equation}
We show the two model response function in Fig.~\ref{fig:models}. The main panel shows the two models $S_A(\omega)$ and $S_B(\omega)$ over the entire frequency range, the presence of the tail in $S_B(\omega)$ is easily seen. In the inset we show instead the two response function in linear scale for the energy range of interest $[-1,-0.8]$, corresponding to the choice $\delta\nu=0.2$. In this range the two responses appear almost identical. In the figure we also report the obtained values for the mean $\mu\equiv\mu_1$ and the square root of the variance $\sigma$ for the two models
\begin{equation}
\begin{split}
\mu^A=-0.911\quad&\quad\sigma^A=0.031\\
\mu^B=-0.907\quad&\quad\sigma^B=0.067\\
\end{split}
\end{equation}
The presence of the high-frequency tail does not modify appreciably the mean value while the variance is increased by more then a factor of two.

Due to the presence of delta function peaks, which are generated by the discreteness of the spectrum of a finite matrix, in general we cannot directly compare the integral transform $\Phi$ with the originating response function $S(\omega)$ since the normalization conventions are different. For the discrete response function we have
\begin{equation}
\int_{-1}^1d\omega S(\omega) = \sum_{k}S(\omega_k)=1\;,
\end{equation}
while for the continuous integral transform instead
\begin{equation}
\int_{-1}^1d\omega \Phi(\omega) = 1\;.
\end{equation}
In order to directly compare the two, we approximate the integral over frequencies with the finite difference
\begin{equation}
\int_{-1}^1d\omega \Phi(\omega) \approx \delta\omega \sum_{n=0}^{2/\delta\omega}\Phi\left(n\delta\omega-1\right)\;,
\end{equation}
and choose $\delta\omega=2/512$ in order to match the average spacing between the eigenvalues. In the following we will then consider the dimensionless quantity $\delta\omega\Phi$ which implies the choice $\Omega=\delta\omega$ for the energy scale of the error. We further choose the the parameters of the integral kernel as
\begin{equation}
\Delta = 0.02\quad\quad\Sigma=0.01\;.
\end{equation}

\begin{figure}[t]
 \resizebox{0.49\textwidth}{!}{ \includegraphics{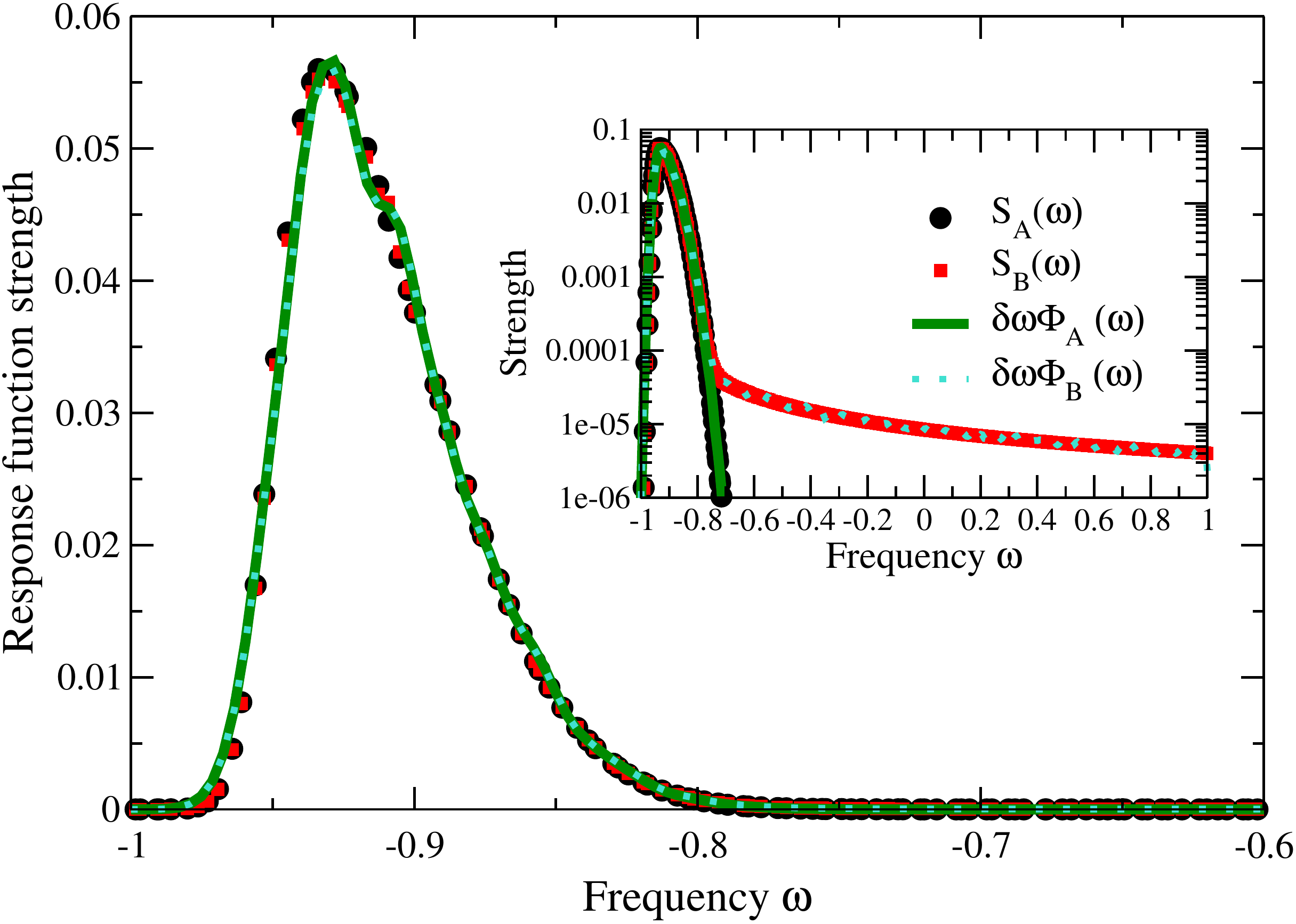} }
  \caption{Comparison between the model response function $S_A(\omega)$ and $S_B(\omega)$ (black dots and red squares respectively) with their rescaled integral transforms $\delta\omega\Phi_A(\omega)$ and $\delta\omega\Phi_B(\omega)$ (green solid line and turquoise dotted line respectively). The inset shows the same curves but over the full energy range. }
  \label{fig:model_it_cmp}
\end{figure}

Fig.~\ref{fig:model_it_cmp} shows a comparison between $S_A(\omega)$ and $S_B(\omega)$ with their rescaled integral transforms $\delta\omega\Phi_A(\omega)$ (shown as a solid green curve) and $\delta\omega\Phi_B(\omega)$ (shown as a dotted turquoise curve). We can clearly see that the rescaling described above allows us to compare them directly.

Since the truncation error $\epsilon_N$ and the sampling error $\epsilon_S$ are more standard and the bounds described in this work are not novel, we concentrate in the following in the analysis of the systematic error $\epsilon_P$ coming from the need to perform a periodic extension of the integral kernel.

\begin{figure}[t]
 \resizebox{0.49\textwidth}{!}{ \includegraphics{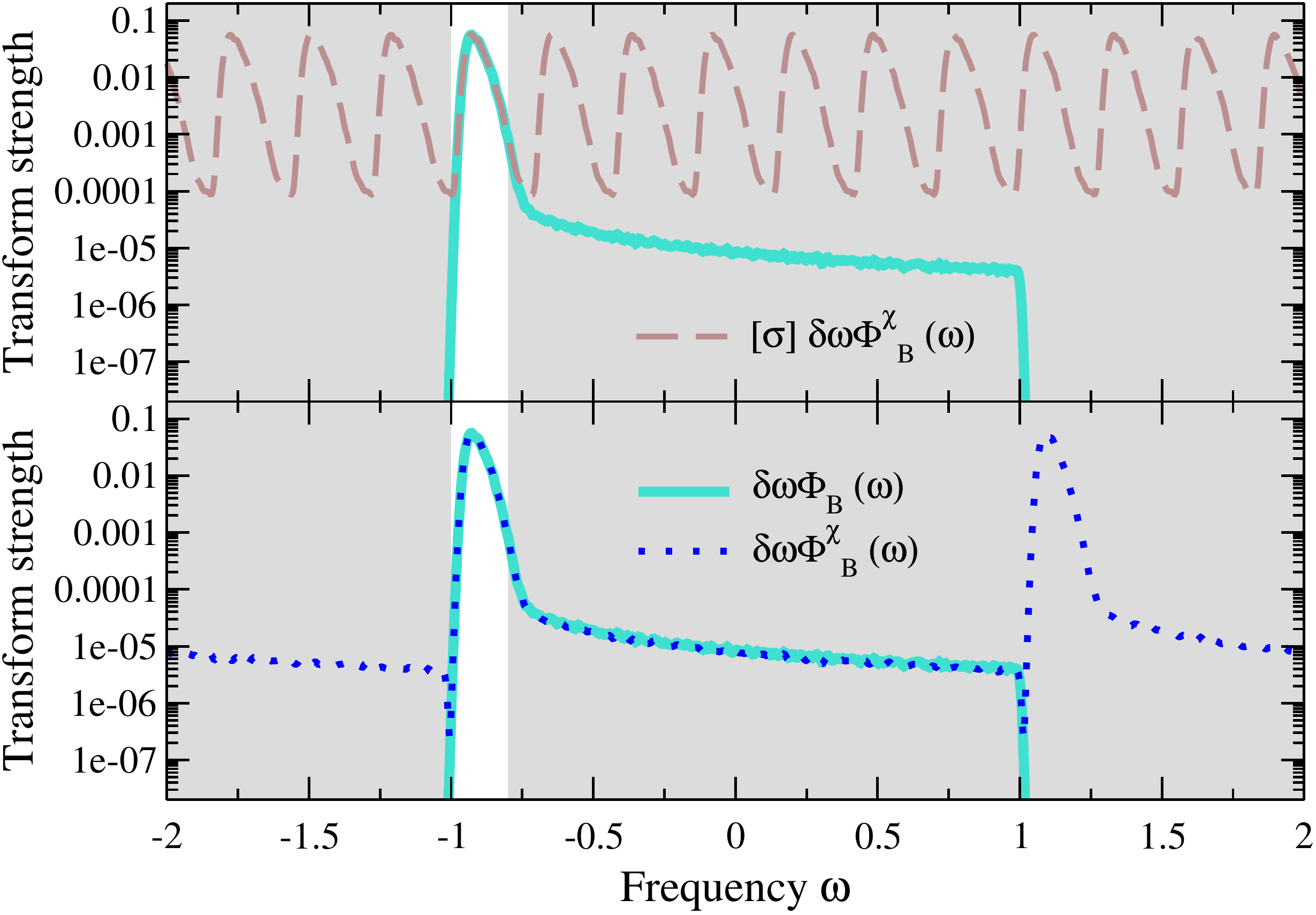} }
  \caption{Comparison between the ideal (scaled) integral transforms $\delta\omega\Phi_B(\omega)$ (turquoise solid lines) and different periodic extensions. The bottom panel shows the approximation obtained using $P$ for a generic response denoted as $\delta\omega\Phi^\chi_B(\omega)$ (blue dotted line) while the top panel shows the approximation obtained with a smaller $P$ found using the known energy variance $\sigma^2$ and denoted as $[\sigma]\;\delta\omega\Phi^\chi_B(\omega)$ (brown solid line). The shaded grey areas are energies outside the interval $[\nu_{\min},\nu_{\max}]$.}
  \label{fig:it_pexp_cmp}
\end{figure}

We present in Fig.~\ref{fig:it_pexp_cmp} a direct comparison between the ideal (scaled) integral transform $\delta\omega\Phi_B(\omega)$ for model $B$ (turquoise solid lines) with the approximations obtained by applying a periodic extension to the integral kernel with different choices of periods $P$ (cf. Eq.~\eqref{eq:period_kernel} above). In all cases we set $\varepsilon_P=0.01$ for the target error. The approximation in the bottom panel, denoted $\delta\omega\Phi_B(\omega)$ (blue dotted line), is obtained using the conservative choice from Eq.~\eqref{eq:chi_gen_main} which is valid in general. We see clearly that the effect is to push the replicas outside the whole range of the Hamiltonian, which in our case is $[-1,1]$, and therefore no appreciable change affects the transform in the required energy range $[\nu_{\min},\nu_{\max}]=[-1,0.8]$: the maximum observed deviation is $\approx10^{-8}$. To highlight the location of this energy range, we have shaded in grey the region of energies outside of it. Also, for ease of visualization, we have presented results in a larger range of energies in order to be able to see where the replicas are located. The top panel shows instead the periodic extensions obtained using the improved bound on $P$ obtained in this work which uses directly prior information about the first and second energy moment. It is apparent that now the integral transform outside the energy window of interest is severely distorted due to the presence of the periodic replicas. However, inside the required region, the maximum deviations are well below the value $\epsilon_P=0.01$ chosen as target: we observe in fact an error $\approx10^{-4}$. Very similar results have been observed for the approximations of the integral transform of model $A$. Before analyzing the relationship between target error and the empirically observed deviation, we want to point out the the reduction in the value of the period are
\begin{equation}
\frac{P^\sigma_A}{P^{gen}}\approx 0.111\quad\quad\frac{P^\sigma_B}{P^{gen}}\approx 0.14\;,
\end{equation}
for the two model responses respectively. Here we denoted by $P^{gen}$ the value obtained without information about the energy moments and with $P^\sigma$ the value obtained using the knowledge of the energy variance. In terms of the required number of moments to guarantee a total approximation error less than $\varepsilon_P+\varepsilon_N=0.02$ we find using Eq.~\eqref{eq:n_bound_main}
\begin{equation}
N^{gen} = 218\quad\quad N_A^{\sigma} = 25\quad\quad N_B^{\sigma} = 31\;.
\end{equation}

\begin{figure}[t]
 \resizebox{0.49\textwidth}{!}{ \includegraphics{error_scaling2.pdf} }
  \caption{The top panel shows, as a function of the target error, the fractional reduction in the period from the genera case value $P=P^{gen}$ obtained using information about the energy variance: the green line shows $P^{\sigma}_A/P$ for model $A$ while the turquoise is $P^{\sigma}_B/P$ for model $B$. The bottom panel shows the observed approximation error for the two models as a function of the target error: the green line for model $A$ and the turquoise line for model $B$. The solid black line represents the best error bound we derived for this approximation (see App.~\ref{sec:pwithmom}).}
  \label{fig:errors}
\end{figure}

In order to better understand the above observation that the measured error seems to be much smaller than the target one, which employs a possibly not tight bound on the possible error, we present in Fig.~\ref{fig:errors} an analysis of the scaling of the empirical error with the one set as target. The bottom panel shows how the approximation error in the scaled integral transform changes as a function of the target error for the cases where we use the improved estimate for the period $P^\sigma$ which uses the energy variance. The solid black line shows the upper-bound employed to control the error while the solid green line and solid turquoise line show the observed maximum errors for both models respectively. We can see that, as expected, the real error is always below the bound we use but that the two models show a very different scaling for small target errors: the integral transform of model $A$ has an error which decreases super-polynomially with target error while for model $B$ the bound is saturated for small enough values of target error. We can understand this behavior in terms of the high-energy tail of the original response functions: as can be seen from the inset of Fig.~\ref{fig:model_it_cmp} the response function for model $B$ is dominated by the slowly decaying tail for energies outside the range of interest $[\nu_{\min},\nu_{\max}]=[-1,-0.8]$ while the strength of model $A$ decays much faster. Once the target error becomes comparable to the strength in the tail of model $B$ then one is forced to consider $P^\sigma= P^{gen}$ as for a general response. On the other hand, for model $A$ as soon as the value of $P^\sigma$ exceeds the range of frequencies of interest than the error decays quickly following the strength response. These results show very clearly the important effect produced by slowly decaying tails in the response function and how their presence is automatically captured by the tail bounds employed in this work. The observation that the error bound is essentially saturated for small errors in model $B$ suggests that it is unlikely to be possible to considerably improve our estimate for the optimal period $P^\sigma$ in general. Finally, for not too small values of the target error $\gtrapprox10^{-3}$ our error bound overestimated the real error by more than an order of magnitude. In this regime it is possible that improved estimates using higher central moments (c.f. Eq.~\eqref{eq:nmom_high} and Eq.~\eqref{eq:high_mu_chi}) would be able to increase the savings in computational cost even further.

\section{Conclusions}
\label{sec:concl}

In this work we have extended the Gaussian Integral Transform method~\cite{git2020} by employing a Fourier basis for the polynomial expansion of the integral trasnform of the spectrl density. This replaces the need to estimate Chebyshev moments of the Hamiltonian with analogous Fourier moments that can be evaluated as expectation values of the real-time evolution operator $\exp(-it\hat{H})$ (c.f. Eq.~\eqref{eq:phi_finite_N}). Importantly, evaluation of both of these moments can be achieved efficiently using existing quantum algorithms. The main advantage of employing a Fourier basis is the possibility of using prior information about the spectral density, in the form of its first few energy moments, to reduce the number of observable needed for an accurate reconstruction of the spectrum. Most of our derivation and numerical examples are focused on the reasonable assumption that only the mean and variance of the energy spectrum are available to the user but we also provide improved bounds in case more moments are available.

The technique presented here can allow for orders of magnitude speed-ups in the evaluation of the spectral density. In order to give some concrete examples in nuclear physics, we take the Hamiltonian derived from Lattice EFT and used in previous cost estimates for quantum computations of the response function (see~\cite{Roggero2020nu}) to model a medium mass nucleus with $A=40$. As a first example we consider a calculation of the dipole response of $^{40}Ca$ for excitation energies up to $100$ MeV, with resolution $\Delta=1$ MeV, small tail contributions $\Sigma=0.01$ and approximation error $\varepsilon=0.01$ (we take $\Omega=\Delta$ as scale). Using the experimental values from Ref.~\cite{AHRENS1975479} (see also~\cite{PhysRevC.90.064619}) we estimate $\mu^{GT}_1\approx20$ MeV and $\sigma^{GT}\approx22$ MeV. With these values we find more than two orders of magnitude reduction in the number of moments required
\begin{equation}
N_{GT}^{gen}=42372\quad\quad N^\sigma_{GT}=339\;.
\end{equation}
As a second example we consider instead the simulation of longitudinal response in quasi-elastic electron-nucleus scattering at momentum transfer $q$. Typical values of the moments are $\mu^{QE}_1=q^2/2M$, with $M$ the nucleon mass, and $\sigma^{QE}\approx k_F\approx250$ MeV, with $k_F$ the Fermi momentum~\cite{PhysRevC.56.3152,PhysRevLett.127.072501}. For a momentum transfer $q=400$ MeV and a maximum excitation energy $\delta\nu=400$ MeV we find
\begin{equation}
N_{QE}^{gen} = 42372\quad\quad N^\sigma_{QE}=838\;.
\end{equation}
An expected saving by a factor $\approx50$.

It is likely that for calculations in this regime the simple low-energy model used in Ref.~\cite{Roggero2020nu} might not be suitable. Due to the expected increase in the Hamiltonian norm, we would expect a larger relative saving in number of moments in cases where higher resolution Hamiltonians with a larger momentum cut-off are employed. 

Lastly we want to comment on the fact that the total computational cost controlled by the maximum evolution time $T=N\delta_t$ still scales as $\mathcal{O}(1/\Delta)$ and we therefore have no violation of the no-fast-forwarding theorem~\cite{Atia2017}. Interestingly for Hamiltonians that can be fast-forwarded~\cite{Atia2017,Gu2021fastforwarding} the computational cost is no longer bounded by $T$ but by $N$ instead. For applications like the reconstruction of the spectral function or the calculation of thermodynamic observables like in Ref.~\cite{Lu2021} this will be completely given by classical cost. One can also employ the construction presented here to prepare states in a given energy window by performing the summation in Eq.~\eqref{eq:kernel_chi_exp} (or more correctly it's finite $N$ approximation) coherently on a quantum device using the Linear Combination of Unitaries strategy~\cite{childs2012}. This is similar to the energy filter proposed in Ref.~\cite{Ge2019} or the original GIT~\cite{git2020} (if used coherently) and the explicit appearance of the energy variance in our cost estimates can prove useful in reducing the cost for some situations. A similar procedure could be employed as a subroutine to the Verified Phase Estimation~\cite{OBrien_2021} in order to estimate general expectation values. We leave a more thorough exploration of these possibilities to future work on spectral filters.

\begin{acknowledgement}
We thank Joseph Carlson for discussions about nuclear response functions. This work was supported in part by the U.S. Department of Energy, Office of Science, Office of Nuclear Physics, Inqubator for Quantum Simulation (IQuS) under Award Number DOE (NP) Award DE-SC0020970
\end{acknowledgement}
\appendix
\section{Derivation of Error Bounds}
\label{sec:error_bounds}
In this appendix we provide a full derivation of the error bounds used in the main text. In order to simplify the notation we will denote the Gaussian kernel as
\begin{equation}
G_\nu(\omega) = \frac{1}{\sqrt{2\pi}\Lambda} \exp\left(-\frac{(\nu-\omega)^2}{2\Lambda^2}\right)\;,
\end{equation}
and it's periodic extension with period P as
\begin{equation}
G^P_\nu(\omega) = \sum_{k=-\infty}^\infty G_\nu(\omega+kP)= \sum_{k=-\infty}^\infty G_{\nu+kP}(\omega)\;.
\end{equation}

We start with the error introduced by using the periodic extension of the kernel
\begin{equation}
\lvert\Phi^\chi(\nu)-\Phi(\nu)\rvert = \epsilon_P(\nu)\;.
\end{equation}

\subsection{Periodic extension in the general case}
\label{ssec:general_chi}

We can write this difference explicitly as
\begin{equation}
\begin{split}
\label{eq:err_P_rewritten}
\epsilon_P(\nu) &= \left|  \int d\omega G^P_\nu(\omega)S(\omega) - \int d\omega G_\nu(\omega)S(\omega) \right|\\ 
&= \left|  \int d\omega\left(\sum_{k=-\infty}^\infty G_{\nu+kP}(\omega) - G_\nu(\omega)\right)S(\omega) \right|\\ 
&= \left|  \int d\omega\sum_{k\neq0} G_{\nu+kP}(\omega)S(\omega) \right|\\ 
&= \int d\omega \sum_{k\neq0} G_{\nu+kP}(\omega)S(\omega) \\ 
\end{split}
\end{equation}
where in the last line we used the fact that both $S(\omega)$ and the Gaussian kernel are positive definite. We can now use the fact that $S(\omega)=0$ for frequencies outside the energy spectrum so that
\begin{equation}
\epsilon_P(\nu) =  \int^{\|\hat{H}\|}_{-\|\hat{H}\|}  d\omega \sum_{k\neq0}G_{\nu+kP}(\omega)S(\omega)\;.
\end{equation}
At this point we can use the fact that $S(\omega)$ is integrable, and in particular we can write
\begin{equation}
\int^{\|\hat{H}\|}_{-\|\hat{H}\|}d\omega S(\omega)=\int d\omega S(\omega) = \mu_0\;,
\end{equation}
the zeroth-moment (cf. Eq.~\eqref{eq:ene_mom} in the main text). Using this result one can then bound the error with
\begin{equation}
\epsilon_P(\nu) \leq \mu_0 \sup_{\omega\in[-\|\hat{H}\|,\|\hat{H}\|]}\quad\sum_{k\neq0}   G_{\nu+kP}(\omega)\;.
\end{equation}

It is convenient to rewrite the summation as follows
\begin{equation*}
\left[G_{\nu+P}(\omega)+G_{\nu-P}(\omega)\right] + \sum_{k=2}^\infty\left[G_{\nu+kP}(\omega)+G_{\nu-kP}(\omega)\right]\;,
\end{equation*}
since the second term can be bounded easily using
\begin{equation}
\sum_{k=2}^\infty G_{\nu+kP}(\omega)\leq \int_1^\infty dx G_{\nu+xP}(\omega)\;,
\end{equation}
\begin{equation}
\begin{split}
\sum_{k=2}^\infty G_{\nu+kP}(\omega)&\leq \int_1^\infty dx G_{\nu+xP}(\omega)\\
&=\frac{1}{\sqrt{2\pi}\Lambda}\int_1^\infty dx e^{-\frac{(\omega-\nu-xP)^2}{2\Lambda^2}}\\
&=\frac{1}{2 P}\text{erfc}\left(\frac{P-\omega+\nu}{\sqrt{2}\Lambda}\right)\\
\end{split}
\end{equation}
so that, including the contribution with $-k$, we find
\begin{equation}
\begin{split}
\sum_{k=2}^\infty \left[G_{\nu+kP}(\omega)+G_{\nu-kP}(\omega)\right]&\leq \frac{1}{2 P}\text{erfc}\left(\frac{P-\omega+\nu}{\sqrt{2}\Lambda}\right)\\
&+\frac{1}{2 P}\text{erfc}\left(\frac{P+\omega-\nu}{\sqrt{2}\Lambda}\right)
\end{split}
\end{equation}

We can now proceed to use the bound in Eq.~\eqref{eq:erfc_bnd} of the main text to write
\begin{equation}
\label{eq:eps_nu_bound}
\epsilon_P(\nu) \leq C\sup_{\omega\in[-\|\hat{H}\|,\|\hat{H}\|]}\left(G_{\nu+P}(\omega)+G_{\nu-P}(\omega)\right)\;,
\end{equation}
where the constant factor $C$ is given by
\begin{equation}
C= \mu_0 \left(1+\sqrt{\frac{\pi}{2}}\frac{\Lambda}{P}\right)\;.    
\end{equation}
In order to turn this into a useful bound, we will consider the largest error that can occur for any 
\begin{equation}
\nu\in[\nu_{\min},\nu_{\max}]\subseteq[-\|\hat{H}\|,\|\hat{H}\|]\;,
\end{equation}
which directly implies that
\begin{equation}
\omega-\nu \in[-\|\hat{H}\|-\nu_{\max},\|\hat{H}\|-\nu_{\min}]\subseteq[-2\|\hat{H}\|,2\|\hat{H}\|]\;.
\end{equation}
First, notice that one of two Gaussians in Eq.~\eqref{eq:eps_nu_bound} will always dominate over the other. To see this we will consider two cases separately: first let's take
\begin{equation}
\omega-\nu \in [-\|\hat{H}\|-\nu_{\max},0]\;.
\end{equation}
In this range of values we have
\begin{equation}
\begin{split}
|P-\omega+\nu|&=P-\omega+\nu\geq P>0\\
|P+\omega-\nu|&=|P-|\omega-\nu||\left\{\begin{matrix}
>0&\text{for}\;\;P>\|\hat{H}\|+\nu_{\max}\\
\geq 0&\text{for}\;\;P\leq\|\hat{H}\|+\nu_{\max}\\
\end{matrix}\right.
\end{split}
\end{equation}
For this range, the second Gaussian centered in $\nu-P$ dominates and we need to have $P>\|\hat{H}\|+\nu_{\max}$ in order to prevent the exponent to go to zero. Therefore
\begin{equation}
\begin{split}
G_{\nu+P}(\omega)+&G_{\nu-P}(\omega)\leq 2G_{\nu-P}(\omega)\\
&\leq \frac{2}{\sqrt{2\pi}\Lambda}\exp\left(-\frac{(P-\|H\|-\nu_{\max})^2}{2\Lambda^2}\right)\;.
\end{split}
\end{equation}
If we take the complementary range of values
\begin{equation}
\omega-\nu \in [0,\|\hat{H}\|-\nu_{\min}]\;,
\end{equation}
we find instead the the first Gaussian dominate and, for $P>\|\hat{H}\|-\nu_{\min}$ we find the useful bound
\begin{equation}
\begin{split}
G_{\nu+P}(\omega)+&G_{\nu-P}(\omega)\leq 2G_{\nu+P}(\omega)\\
&\leq \frac{2}{\sqrt{2\pi}\Lambda}\exp\left(-\frac{(P-\|H\|+\nu_{\min})^2}{2\Lambda^2}\right)\;.
\end{split}
\end{equation}
These results suggest that we should take
\begin{equation}
P=(1+\eta) \|\hat{H}\|+\max\left[|\nu_{\min}|,\nu_{\max}\right]\;,
\end{equation}
for some appropriate $\eta>0$. For this choice we have in fact
\begin{equation}
\epsilon_P(\nu) \leq \frac{2\mu_0}{\sqrt{2\pi}\Lambda}\left(1+\sqrt{\frac{\pi}{2}}\frac{\Lambda}{P}\right)e^{-\frac{\eta^2\|\hat{H}\|^2}{2\Lambda^2}}\;.
\end{equation}
In order to simplify the calculation of a good value for $\eta$ that would guarantee $\epsilon_P(\nu)\Omega<\varepsilon_P$, for some (dimensionless) target error tolerance $\varepsilon_P>0$, we use the simple lower bound $P>\|\hat{H}\|$ and find
\begin{equation}
\label{eq:etabound}
\eta\geq \frac{\sqrt{2}\Lambda}{\|\hat{H}\|}\sqrt{\log\left(\sqrt{\frac{2}{\pi}}\frac{\mu_0}{\varepsilon_P}\frac{\Omega}{\Lambda}\left(1+\sqrt{\frac{\pi}{2}}\frac{\Lambda}{\|\hat{H}\|}\right)\right)}\;.
\end{equation}
Neglecting sub-leading logarithmic factors we find therefore the asymptotic scaling
\begin{equation}
\chi=\frac{P}{\|\hat{H}\|}=\widetilde{\mathcal{O}}\left(1+\frac{\Lambda}{\|\hat{H}\|}\sqrt{\log\left(\frac{\mu_0\Omega}{\varepsilon_P\Lambda}\right)}\right)\;.
\end{equation}

A more practical bound which uses the conventions of the main text is the following one
\begin{equation}
\chi=2+\frac{\sqrt{2}\Lambda}{\|\hat{H}\|}\sqrt{\log\left(\frac{2\Omega}{\varepsilon_P\Lambda}\right)}\;,
\end{equation}
which is the result quoted in the main text.

\subsection{Periodic extension with moment information}
\label{sec:pwithmom}

We now turn to show how to obtain an improved error bound using information about the energy moments. As before, and without loss of generality, we will consider $\nu\in[\nu_{\min},\nu_{\max}]$. For the following manipulations, it will be convenient to change variables and define
\begin{equation}
\nu_{\min}=\mu_1-\Omega_{\min}\quad\nu_{\max}=\mu_1+\Omega_{\max}\;,
\end{equation}
with $\mu_1$ the first energy moment which gives us information about the average energy in the spectral function and $\Omega_{\min}$,$\Omega_{\max}$ both positive. At this point we start from the expression for the error $\epsilon_P(\nu)$ obtained in Eq.~\eqref{eq:err_P_rewritten} before and rewrite it as follows
\begin{equation}
\begin{split}
\epsilon_P(\nu)=&\int d\omega \sum_{k\neq0}G_{\nu+kP}(\omega)S(\omega)\\
=&\int_{-\infty}^{\mu_1-\alpha\sigma} d\omega \sum_{k\neq0}G_{\nu+kP}(\omega)S(\omega)\\
&+\int_{\mu_1-\alpha\sigma}^{\mu_1+\alpha\sigma} d\omega \sum_{k\neq0}G_{\nu+kP}(\omega)S(\omega)\\
&+\int_{\mu_1+\alpha\sigma}^{\infty} d\omega \sum_{k\neq0}G_{\nu+kP}(\omega)S(\omega)\\
=&\epsilon_1(\nu)+\epsilon_2(\nu)+\epsilon_3(\nu)\;,
\end{split}
\end{equation}
for some $\alpha>0$ and where we have denoted by $\sigma$ the square root of the variance
\begin{equation}
\sigma = \sqrt{\mu_2-\mu_1^2}\;.
\end{equation}
The central integral can be bounded in the same way we obtained the result in the previous section, the main difference is that now we take the interval
\begin{equation}
\omega-\nu\in\left[-\alpha\sigma-\Omega_{\max},\alpha\sigma+\Omega_{min} \right]\;.
\end{equation}
In this range we can bound the central contribution with
\begin{equation}
\epsilon_2(\nu)\leq \frac{2\mu_0}{\sqrt{2\pi}\Lambda}\left(1+\sqrt{\frac{\pi}{2}}\frac{\Lambda}{P}\right)e^{-\frac{\eta^2\alpha^2\sigma^2}{2\Lambda^2}}\;,
\end{equation}
where now we took the period to be
\begin{equation}
\label{eq:pchoice}
P=(1+\eta)\alpha\sigma+\max\left[\Omega_{\min},\Omega_{\max}\right]\;,
\end{equation}
for some $\eta>0$. Note that, in order for the exponential to become small, we would need $P>\eta\alpha\sigma>\Lambda$ so that we can use the simpler bound
\begin{equation}
\epsilon_2(\nu)\leq \frac{2\mu_0}{\sqrt{2\pi}\Lambda}\left(1+\sqrt{\frac{\pi}{2}}\right)e^{-\frac{\eta^2\alpha^2\sigma^2}{2\Lambda^2}}\;,
\end{equation}
which will incur at most a logarithmic cost. In order to guarantee this term to be smaller than $\varepsilon_P/(2\Omega)$, for dimensionless $\varepsilon_P>0$, we can take (cf. Eq.~\eqref{eq:etabound})
\begin{equation}
\label{eq:eta_choice}
\eta=\frac{\sqrt{2}\Lambda}{\alpha\sigma}\sqrt{\log\left(\sqrt{\frac{8}{\pi}}\frac{\mu_0}{\varepsilon_P}\frac{\Omega}{\Lambda}\left(1+\sqrt{\frac{\pi}{2}}\right)\right)}\;.
\end{equation}

In order to bound the other two terms, we first bound the sum over $k$ with an integral using
\begin{equation}
\sum_{k=1}^\infty G_{\nu+kP}(\omega)\leq \int_0^\infty dx G_{\nu+xP}(\omega)\;,
\end{equation}
together with a similar one for the negative terms
\begin{equation}
\begin{split}
\sum_{k=1}^\infty G_{\nu-kP}(\omega)&\leq \int_0^\infty dx G_{\nu-xP}(\omega)\\
&=\int_{-\infty}^0 dx G_{\nu+xP}(\omega)\;,
\end{split}
\end{equation}
so that summing them together we find the bound
\begin{equation}
\sum_{k\neq0}^\infty G_{\nu+kP}(\omega)\leq\int_{-\infty}^\infty dx G_{\nu+xP}(\omega)=\frac{1}{P}\;.
\end{equation}
The sum of the two missing terms $\epsilon_{13}(\nu)=\epsilon_1(\nu)+\epsilon_3(\nu)$ is thus bounded by the following
\begin{equation}
\begin{split}
\epsilon_{13}(\nu)&\leq\frac{1}{P}\left(\int_{-\infty}^{\mu-\alpha\sigma}d\omega S(\omega)+\int_{\mu+\alpha\sigma}^\infty d\omega S(\omega)\right)\\
&=\frac{1}{P}\left(1-\int_{\mu-\alpha\sigma}^{\mu+\alpha\sigma}d\omega S(\omega)\right)\\
&\leq \frac{\sigma^2}{P\alpha^2\sigma^2}=\frac{1}{P\alpha^2}\\
&\leq \frac{1}{(1+\eta)\alpha^3\sigma}\;.
\end{split}
\end{equation}
Here we used the normalization of $S(\omega)$ to get to the second line, the Chebyshev bound Eq.~\eqref{eq:Cheb_bound} in the main text to get to the third and Eq.~\eqref{eq:pchoice} for the last one. At this point we can in principle use the expression for $\eta$ in Eq.~\eqref{eq:eta_choice} to find an appropriate value for $\alpha$ so that $\epsilon_{13}(\nu)<\varepsilon_P/(2\Omega)$. In order to better see the general trend, we instead ensure that $\alpha$ satisfies
\begin{equation}
\alpha\geq\frac{\sqrt{2}\Lambda}{\sigma}\sqrt{\log\left(3.6\frac{\mu_0\Omega}{\varepsilon_P\Lambda}\right)}\;,
\end{equation}
so that $\eta<1$. We can then solve
\begin{equation}
\frac{1}{(1+\eta)\alpha^3\sigma}\leq\frac{1}{\alpha^3\sigma}\leq\frac{\varepsilon_P}{2\Omega}\;.
\end{equation}
This results in the following bound
\begin{equation}
\label{eq:optalpha}
\alpha\geq \frac{\sqrt{2}\Lambda}{\sigma}\max\left[\sqrt{\log\left(3.6\frac{\mu_0\Omega}{\varepsilon_P\Lambda}\right)},\frac{\sigma^{2/3}\Omega^{1/3}}{\Lambda}\frac{0.9}{\varepsilon_P^{1/3}}\right]\;.
\end{equation}
This result can be inserted directly into Eq.~\eqref{eq:pchoice} to obtain the final estimate for the period $P$ required.

We now use the conventions described in the main text to obtain a more intuitive result and distinguish explicitly two different regimes. For $2\sigma\geq\Lambda$ the second term will dominate for all $\varepsilon_P\lesssim0.6$ and for all values if we increase the numerical factor to $1.9$. In this case we can take
\begin{equation}
\chi=  \frac{2.7}{\varepsilon_P^{1/3}} \frac{\Omega^{1/3}\sigma^{2/3}}{\|\hat{H}\|}+\frac{\delta\nu}{\|\hat{H}\|}\;,
\end{equation}
as quoted in the main text. Here we used the fact that
\begin{equation}
\max\left[\Omega_{\min},\Omega_{\max}\right]\leq\Omega_{\min}+\Omega_{\max}=\nu_{\max}-\nu_{\min}=\delta\nu\;.
\end{equation}
For large values of $\Lambda>2\sigma$ we would need to use the full expression in Eq.~\eqref{eq:optalpha} above instead.

Finally, for cases where we know the value of central moments 
\begin{equation}
\widetilde{\mu}_n = \int d\omega \left|\omega-\mu_1\right|^nS(\omega)\;,
\end{equation}
for $n$ higher then $2$ are known, then one can use a the following generalization of the Chebyshev inequality
\begin{equation}
Prob\left[\left|\omega-\mu_1\right|\geq\Gamma\right]\leq\frac{\widetilde{\mu_n}}{\Gamma^n}\;.
\end{equation}
Using this additional information we can take instead
\begin{equation}
\label{eq:pnchoice}
P=(1+\eta)\alpha\widetilde{\mu_n}^{1/n}+\max\left[\Omega_{\min},\Omega_{\max}\right]\;,
\end{equation}
and can therefor choose $\alpha$ as follows
\begin{equation}
\alpha\geq \frac{\sqrt{2}\Lambda}{\widetilde{\mu_n}^{\frac{1}{n}}}\max\left[\sqrt{\log\left(4\frac{\mu_0\Omega}{\varepsilon_P\Lambda}\right)},\left(\frac{\widetilde{\mu_n}\Omega}{\varepsilon_P}\right)^{\frac{1}{n+1}}\frac{0.9}{\Lambda}\right]\;.
\end{equation}
For $\widetilde{\mu_n^{1/n}}\geq \Lambda$ we have the following simpler result
\begin{equation}
\label{eq:high_mu_chi}
\chi=  \frac{2.7}{\varepsilon_P^{1/(n+1)}} \frac{(\Omega\widetilde{\mu_n})^{1/(n+1)}}{\|\hat{H}\|}+\frac{\delta\nu}{\|\hat{H}\|}\;,
\end{equation}
valid up to $n=15$. This can be advantageous for small errors provided the central moments do not grow too much.

\subsection{Truncation error}
\label{sec:epsilon_N}
Here we provide the details underlying the bound for the truncation error
\begin{equation}
\epsilon_N(\nu)=\lvert\Phi^\chi_N(\nu)-\Phi^\chi(\nu)\rvert\;,
\end{equation}
used in the main text. In order to simplify the notation we will use $m^\chi_n$ to denote the moments (see Eq.~\eqref{eq:moments}). For our Fourier polynomials these are
\begin{equation}
\begin{split}
m^\chi_n &= \int d\omega S(\omega) \exp\left(-i\frac{2\pi}{\chi\|\hat{H}\|}n\omega\right)\\
&=\langle\Psi_0\lvert\hat{O}^\dagger \exp\left(-i\frac{2\pi}{\chi\|\hat{H}\|}n\hat{H}\right)\hat{O}\rvert\Psi_0\rangle\;,
\end{split}
\end{equation}
In addition, note that $|m^\chi_n|\leq\mu_0$ with $\mu_0$ the zeroth energy moment. This can be easily seen by normalizing the state $\hat{O}\rvert\Psi_0\rangle$ with $\sqrt{\mu_0}$ and using the fact that the evolution operator is unitary. We can now express explicitly the truncation as follows
\begin{equation}
\epsilon_N(\nu)=\frac{1}{\chi\|\hat{H}\|}\left|\sum_{n=N+1}^\infty\left(g_n^\chi(\nu)m^\chi_n+g_{-n}^\chi(\nu)m^\chi_{-n}\right)\right|\;,
\end{equation}
with $g^\chi_n(\nu)$ the Fourier coefficients from Eq.~\eqref{eq:fcoeff}. Using the bound on the moments $m^\chi_n$ we then find immediately
\begin{equation}
\begin{split}
\label{eq:en_bound_der}
\epsilon_N(\nu)&\leq\frac{\mu_0}{\chi\|\hat{H}\|}\sum_{n=N+1}^\infty\left(\left|g_n^\chi(\nu)\right|+\left|g_{-n}^\chi(\nu)\right|\right)\\
&=\frac{2\mu_0}{\chi\|\hat{H}\|}\sum_{n=N+1}^\infty\exp\left(-\frac{2\pi^2\Lambda^2}{\chi^2\|\hat{H}\|^2}n^2\right)\\
&\leq\frac{2\mu_0}{\chi\|\hat{H}\|}\int_{N}^\infty dx\exp\left(-\frac{2\pi^2\Lambda^2}{\chi^2\|\hat{H}\|^2}x^2\right)\\
&=\frac{\mu_0}{\sqrt{2\pi}\Lambda}\text{erfc}\left(N\frac{\sqrt{2}\pi\Lambda}{\chi\|\hat{H}\|}\right)\\
&\leq \frac{\mu_0}{\sqrt{2\pi}\Lambda}\exp\left(-N^2\frac{2\pi^2\Lambda^2}{\chi^2\|\hat{H}\|^2}\right)
\end{split}
\end{equation}
where we used the triangle inequality on the first line, the bound on a sum with an integral in the third and Eq.~\eqref{eq:erfc_bnd} in the last. Note that the bound does not depend on the frequency $\nu$ anymore. We can now find the value for $N$ that would guarantee $\epsilon_N(\nu)\Omega\leq\varepsilon_N$ for some (dimensionless) target error tolerance $\varepsilon_N>0$. The result is
\begin{equation}
\label{eq:nbound_app}
N\geq \frac{\chi\|\hat{H}\|}{\sqrt{2}\pi\Lambda}\sqrt{\log\left(\frac{\mu_0\Omega}{\sqrt{2\pi}\Lambda\varepsilon_N}\right)}\;.
\end{equation}

This is the result used in the main text.

\subsection{Statistical error}
\label{sec:stat_err_app}

We now turn to the discussion of the bound on the statistical error in the evaluation of the Fourier moments $m^\chi_n$. For the moment we will assume we have estimated each moment with a fixed error $\varepsilon_M$ which for simplicity we take to be equal for all moments (thanks to the rapid decrease in the coefficients an adaptive strategy be more advantageous). We also neglect the error on the zeroth moment $m^\chi_0$ since its value is known beforehand. Since errors on different moments are independent, we add the error contributions in quadrature to find
\begin{equation}
\begin{split}
\epsilon_S^2&\leq \frac{\varepsilon_M^2\mu_0^2}{\chi^2\|\hat{H}\|^2}\sum_{n=1}^N\left(\left|g_n^\chi(\nu)\right|^2+\left|g_{-n}^\chi(\nu)\right|^2\right)\\
&= 2\frac{\varepsilon_M^2\mu_0^2}{\chi^2\|\hat{H}\|^2}\sum_{n=1}^N\exp\left(-\frac{4\pi^2\Lambda^2}{\chi^2\|\hat{H}\|^2}n^2\right)\\
&\leq 2\frac{\varepsilon_M^2\mu_0^2}{\chi^2\|\hat{H}\|^2}\int_0^N dx\exp\left(-\frac{4\pi^2\Lambda^2}{\chi^2\|\hat{H}\|^2}x^2\right)\\
&< 2\frac{\varepsilon_M^2\mu_0^2}{\chi^2\|\hat{H}\|^2}\int_0^\infty dx\exp\left(-\frac{4\pi^2\Lambda^2}{\chi^2\|\hat{H}\|^2}x^2\right)\\
&=\frac{\varepsilon_M^2\mu_0^2}{2\pi\chi\|\hat{H}\|\Lambda}\;,
\end{split}
\end{equation}
where we used the fact that the variance of the moments is less than $\mu^2_0$ and the same procedure employed in Eq.~\eqref{eq:en_bound_der}. 
In order to attain a total expected error $\epsilon_S\leq\varepsilon_S/\Omega$ we then need to take
\begin{equation}
\varepsilon_M< \frac{\varepsilon_S}{\mu_0}\frac{\sqrt{2\pi\chi\|\hat{H}\|\Lambda}}{\Omega}
\end{equation}
resulting in an expected number of samples
\begin{equation}
S> 2N\frac{\Omega^2\mu_0^2}{2\pi\chi\|\hat{H}\|\Lambda\varepsilon_S^2}\;.
\end{equation}
The additional factor of 2 in the numerator comes from the need to evauate separately the real and imaginary part of each moment separately.
If we want to ensure this is sufficient with high probability we can use Markov's to ensure the probability that the error is below $\epsilon_S^2$ is larger than $2/3$ by increasing the target error $\varepsilon_S$ by a factor of at least $\sqrt{3}$ and then use the Chernoff bound and majority voting to increase the probability to $1-\delta$ with logarithmic effort. For instance
\begin{equation}
S=N\frac{\Omega^2\mu_0^2}{\chi\|\hat{H}\|\Lambda\varepsilon_S^2}\log\left(\frac{2}{\delta}\right)\;,
\end{equation}
will be enough for a confidence level $1-\delta$. Together with the bound from Eq.~\eqref{eq:nbound_app} this shows that the number of samples is independent from the number of terms $N$.

The treatment above assumes errors are completely uncorrelated which might not necessarily be the case due to the need of controlling systematic errors with e.g. error mitigation techniques. For a more conservative error estimate we consider instead a bound to $\epsilon_S$ obtained by summing the individual errors in absolute value. Following the same procedure used above we find the final result
\begin{equation}
S=N\frac{\Omega^2\mu_0^2}{\Lambda^2\varepsilon_S^2}\log\left(\frac{2}{\delta}\right)\;,
\end{equation}
quoted in the  main text. We want to conclude this appendix with a similar result regarding the original Chebyshev based GIT from Ref.~\cite{git2020}. The estimate of the sample complexity reported there didn't use the strategy employed here and as a result the original work gave an estimate $S=\mathcal{O}(N^3)$ which was somewhat pessimistic. Using in fact the present strategy, together with the results in App.D.3 of~\cite{git2020} and restoring the energy dimensions in order to be compatible with our current conventions we can show that a number of samples given by
\begin{equation}
S_{Cheb}=2N\frac{\Omega^2}{\Lambda^2\varepsilon_S^2}\log\left(\frac{2}{\delta}\right)\;,
\end{equation}
are enough to control the statistical errors.

\bibliographystyle{unsrt}
\bibliography{refs}
\end{document}